\documentclass[aps,prb,twocolumn,10pt,showpacs]{revtex4-1}

\usepackage{amsmath}
\usepackage{graphicx}
\usepackage{times}
\usepackage{amssymb}
\usepackage{hyperref}
\usepackage{textcomp}
\usepackage{xcolor}
\usepackage[T1]{fontenc}

\begin{document}
\title{Superfluorescence spectra of excitons in quantum wells}
\author{P. Gr\"unwald}
\email{Electronic address: peter.gruenwald2@uni-rostock.de}
\author{G. K. G. Burau}
\author{H. Stolz}
\author{W. Vogel}
\affiliation{Institut f\"ur Physik, Universit\"at Rostock, D-18055 Rostock, Germany}
\date{\today}

\begin{abstract}
We study the fluorescence light emitted from GaAs excitons in semiconductor quantum wells. The excitons are modeled as interacting bosons. By combining quantum optical methods for the excitonic emission spectrum with many particle descriptions of the transmission through the medium, we can evaluate the spectra outside the well. Comparing with experimental spectra, we get a very good agreement. The method helps explaining the main features of the observed spectra. It is demonstrated that the observed spectra show clear evidence of superfluorescent emission.
\end{abstract}


\maketitle

\section{Introduction}\label{sec.intro}

The theoretical description of the emission of coherently excited semiconductor materials on a consistent, quantum optical basis is a long-standing, daunting task. From the many body point of view the problem is to solve the coupled semiconductor Bloch equations, which lead to an infinite hierarchy of coupled equations. These equations are then truncated using cluster expansion techniques~\cite{Kira1}, see e.g.~\cite{Koch1} for luminescence from a quantum dot and~\cite{KochStolz} for luminescence of a quantum well. However, this method cuts off parts of the correlations, which are required to describe the quantum state of the light. Another established ansatz is to apply nonequilibrium Greens functions to solve the wave equations for the fields within the medium. This has proven efficient to compute the susceptibility (optical response) of the medium for excitation strengths up to the Mott-transition, corresponding well to experimental results~\cite{Manzke,Burau-1}. However, the emission spectra were not 
calculated within this approach.

In experiments on resonant Rayleigh scattering (RRS), the emission field of a semiconductor after resonant excitation has been studied~\cite{Biex,Hegarty,Stolz1,Stolz2,Savona}. There has also been considerable work on RRS of the speckle structure of localized excitons in quantum wells, see~\cite{Langbein1}. More recently, a long-distance interaction between exciton spots was demonstrated~\cite{Langbein2}.
However, for all these cases pumping was assumed very weak, and hence, only the coherent emission was considered. The incoherent emission spectra of the excitons were neglected. For the description of quantum optical properties of the fluorescent field, the details about the incoherent emission are necessary. We will thus focus on the incoherent emission, in particular the emission spectra.

An approach to combine the two descriptions of quantum fields and many body effects has led to a consistent formulation for the transmission of quantum fields through an excited material~\cite{Dima1}. An input-output formalism was used to describe a scenario where light fields propagate through a preexcited slab geometry. The output field was shown to consist of a transmitted part from the input fields and the spontaneous emission of the excitons inside the material.
However, a full quantum optical description of the fields originating from these quasiparticles is a persisting problem.

From a quantum optical point of view, a model Hamiltonian for the internal excitations, the Wannier-excitons, has to be proclaimed, which is solved to obtain the quantum light fields from such systems. For very low densities, excitons behave like bosons~\cite{Haug}, while in lowest order exciton-exciton interaction can be described by a Kerr-nonlinearity~\cite{Hana}.
On the other hand, localized excitons within quantum dots act like two-level atoms, emitting sub-Poisson light~\cite{Michler,MichlerExp} and showing strong light-matter coupling~\cite{Shih-1,Shih-2}. Hence, there is a physical motivation for the model Hamiltonian. Yet, it does not directly include other electrons and holes and thus their effects on the light fields.

In this contribution we combine the concepts of quantum optical emission with the medium effects to interpret the fluorescence spectra of an excited GaAs quantum well which has been studied experimentally~\cite{Burau-1}.
We consider a bosonic multi-exciton model including exciton-exciton interaction described with a Kerr-nonlinearity~\cite{Hana}. For multiple excitons in a small region as in the given experiments, we show the analogy to the single exciton scenario. Using a simple model for the response of the medium and computing the optical emission, we obtain resonance fluorescence spectra matching the main features of the experiments: the asymmetric shape of the incoherent emission; the magnitude of the resonance shift; the decrease of the Rayleigh peak for increasing laser intensity below the Mott-transition. 
Another important result is, that the emitted light is in a superfluorescent steady state. Superfluorescence light is obtained from a system of multiple emitters, which couple coherently to the same mode of light~\cite{Dicke,Bonifacio}. Thus the emitters produce a signal field, much more intense than the sum of a the single emitter fields. Originally described for ensembles of atoms, it was later also predicted and shown in exciton systems, see~\cite{Hana-book}. Likewise for very dense systems of excitons, cooperative emission under strong magnetic fields was observed~\cite{Kono}. There is also steady-state superfluorescence~\cite{Bonifacio2}, where the pump laser drives the ensemble into a collective, superradiative state. We develop a criterion for this kind of superfluorescent emission, apply it to the quantum well spectra and obtain clear evidence of superfluorescence.

The paper is organized as follows. In Sec.~\ref{sec.exp}, we describe the experiments that were performed on GaAs quantum wells. From the results of these experiments, we derive in Sec.~\ref{sec.Ham} an effective Hamiltonian for the exciton system. Afterwords, in Sec.~\ref{sec.verify}, we develop criteria to estimate the degree of superfluorescent emission. The output spectra of the quantum well will be analyzed in Sec.~\ref{sec.in-out} for different excitations. Based on our theory together with the experimental results, in Sec.~\ref{sec.super} we provide evidence of the superfluorescent emission. Finally, in Sec.~\ref{sec.con} we give some conclusions and an outlook.

\section{Experiments}\label{sec.exp}
Let us start with a short discussion of the performed experiments in order to formulate a model Hamiltonian for the excitons. More details are given in~\cite{Burau-1,Burau-3}. The studied structure was created via molecular beam epitaxy. It incorporates multiple stacked GaAs quantum wells with different well thicknesses and alternating AlAs-GaAs layers in between to separate the active quantum films~\cite{Schwedt}. It was illuminated with a tunable single mode semiconductor cw laser (linewidth 1 MHz) source from a tilted angle. This allowed observing the emitted light with very high spatial resolution (FWHM = 350 nm). Resonance fluorescence emissions of different excitonic structures were analyzed both spatially and spectrally. The specially designed detection system covers a large solid angle of the surroundings of the probe and allows to collect a large portion of the fluorescence light. 

\begin{figure}[h]
    \includegraphics[width=8cm]{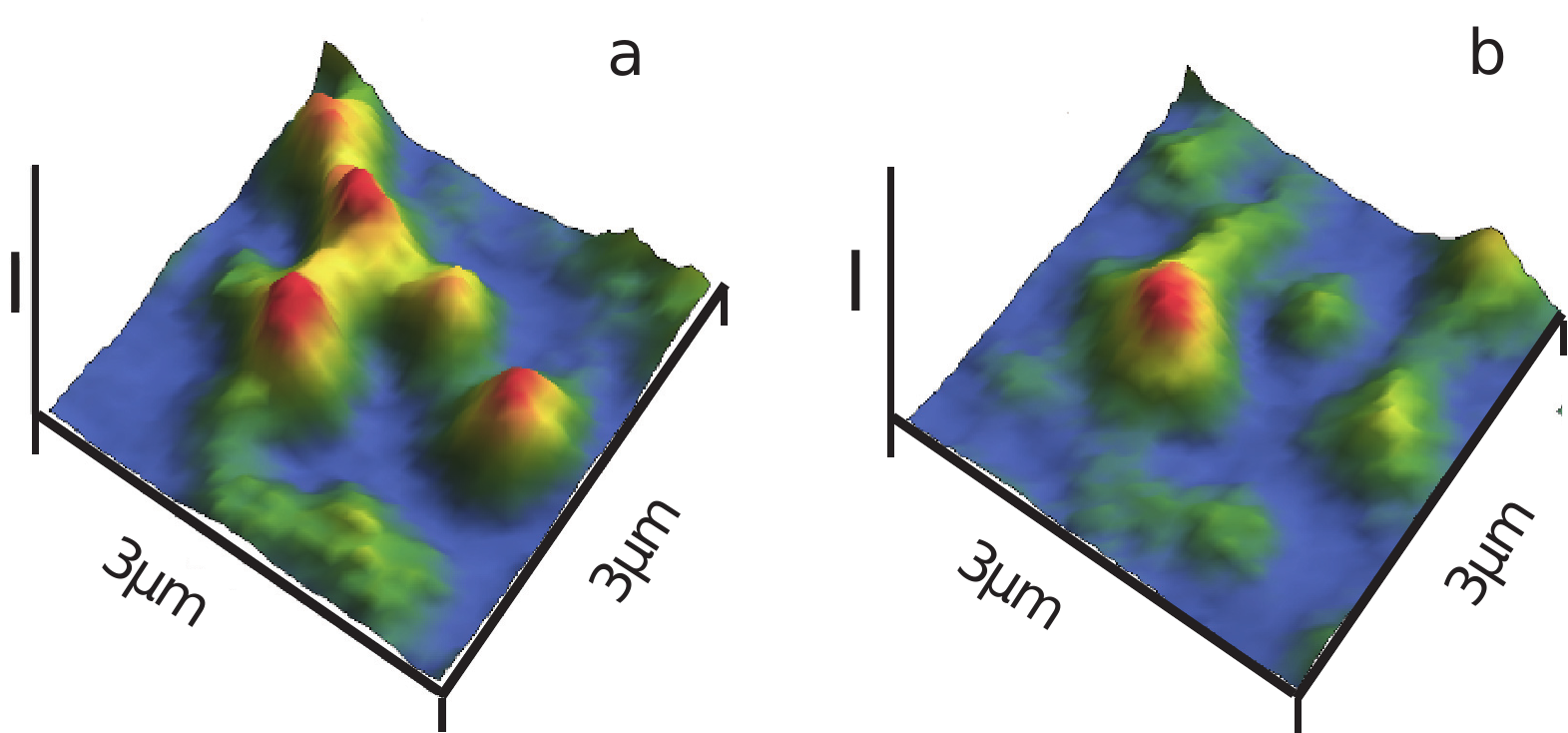}
    \caption{(color online.) Experimental example of the spatially resolved fluorescence intensity $I$ showing different ESs at $T=5$ K for different optical frequencies $\hbar\omega=\hbar \omega_\text L+\Delta$. a: $\Delta=0$, b: $\Delta=25.6$ $\mu$eV.}\label{fig.spatres}
\end{figure}

While multiple layers were analyzed, we will focus on one with a width of 19.8 nm, which was thoroughly studied in terms of the spatio-spectral properties. 
The spatially resolved fluorescence intensity has been recorded, see Fig~\ref{fig.spatres}.
Distributed over the large laser spot, we find localized excitonic structures, so called exciton spots (ESs). The localization of the ESs stems from the surface roughness of the well~\cite{Schwedt,Runge}; hence it is reproducible. With increasing temperature, these ESs grow and combine to form larger spots. Because of the limit in spatial resolution, the number of excitons within one such spot cannot be determined with certainty.

\begin{figure}[h]
  \includegraphics[width=8cm]{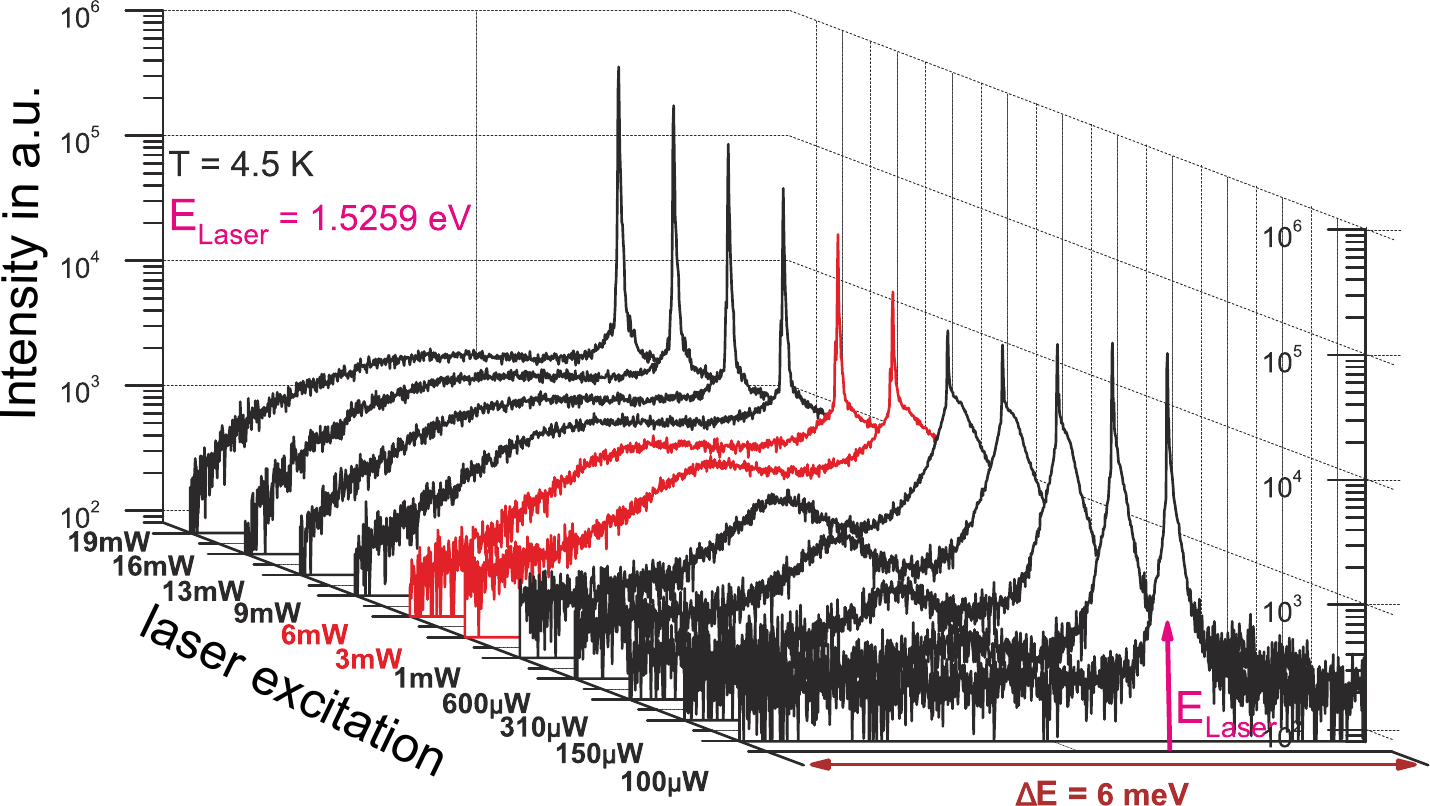}
  \caption{(color online.) Optical fluorescence spectra for one ES and different excitation strengths. The bright (red) lines indicate the Mott-transition, after~\cite{Burau-1}.}\label{fig.ExpRF}
\end{figure}

Single ESs could be analyzed using a pinhole.
Each ES behaves differently (resonance frequency, linewidth, intensity etc.), whereas within one ES, all excitons acquire a near identical state. Notwithstanding this, the qualitative behavior of different spots is observed to be the same.
The optical spectra of the emitted light fields around the 1s resonance of the heavy-hole exciton of a single ES  are depicted for different laser intensities in Fig.~\ref{fig.ExpRF} and discussed below.

For low laser intensities, additionally to a sharp Rayleigh peak at the laser frequency, a homogeneously broadened incoherent spectrum was observed. An ideal bosonic structure of the excitons only yields the Rayleigh peak, not an incoherent spectrum, when irradiated with coherent light. Hence, there must be a nonlinear contribution due to exciton-exciton interaction. However, in difference to two-level atoms~\cite{Stroud} and quantum dots ~\cite{Shih-1,Shih-2}, no Mollow-triplet arises. Thus the excitons also do not behave like atoms in the case under study.

Increasing the intensity, the exciton resonance shifts to a different frequency, with the direction of the shift depending on the sample temperature. 
This shift was discussed in~\cite{Burau-1,Burau-3} and interpreted on the basis of screening effects. It can be described within a model for the response of the medium via its susceptibility $\chi(\omega)$. At the same time, the Rayleigh peak decreases with increasing intensity. An observed sideband at lower frequencies is assumed to be due to the biexciton~\cite{Biex,Biex2}. For high pump intensities, the Mott-transition occurs, which is the main subject of the discussions in~\cite{Burau-3,Manzke}. Above the transition we only see a broad continuum due to the scattering of the electron-hole-plasma, while the Rayleigh peak now increases again.

\section{Exciton model}\label{sec.Ham}
With these results in mind we will now formulate a model Hamiltonian as well as a master equation to describe the emitted fields for lower excitations (below the Mott-transition).
As we work in the low density limit, we consider the excitons within one ES as bosonic. Based on the assumption of identical excitons within one ES, we use the following reasonable approximations. All $N$ excitons of the spot are localized in a small spectral and spatial region, thus all have the same bare transition frequency $\omega_\text x$, where already many-body induced shifts are supposed to be included. Due to localization they have a negligible momentum. The exciting laser spot is much larger than the ES size, such that every exciton is coupled to the laser field with the same Rabi-frequency $\Omega_\text R$. Likewise, also the Kerr-nonlinearity, describing the exciton-exciton interaction, should have the same strength $G$~\cite{Yama,Ciuti,SpFe}.

The excitons are described by bosonic creation and annihilation operators $\hat a^\dagger_n$ and $\hat a_n$, respectively, and are driven by a cw-laser of frequency $\omega_\text L=\omega_\text x-\delta$. The Hamiltonian in the frame of $\omega_\text L$ reads as
\begin{equation}
\begin{split}
    \hat H=&\hbar\sum_n\left\{\delta\hat a^\dagger_n\hat a_n+\Omega_\text R\hat a_ne^{i\phi_n}+\Omega_\text R^*\hat a^\dagger_ne^{-i\phi_n}\right\}\\
	   &+\hbar G\sum_{n,k}\hat a^\dagger_n\hat a^\dagger_k\hat a_k\hat a_n.\label{eq.Multi-Ham}
\end{split}
\end{equation}
Finally, each exciton emits light with the same emission rate $\Gamma$. Hence the full master equation for the density operator $\hat\varrho$ reads as
\begin{equation}
\begin{split}
  \dot{\hat\varrho}&=\frac{1}{i\hbar}[\hat H,\hat\varrho]+\frac{\Gamma}{2}\mathcal L(\hat\varrho)\\
\mathcal L(\hat\varrho)&=\sum_n(2\hat a_n\hat\varrho\hat a^\dagger_n-\hat a^\dagger_n\hat a_n\hat\varrho-\hat\varrho\hat a^\dagger_n\hat a_n).\label{eq.Master}
\end{split}
\end{equation}
The second term is a standard form for the radiative decay~\cite{vowe}, here applied to the excitons. In the following, we will transform this Hamiltonian into an effective Hamiltonian describing a collective $N$-exciton-state.

As the size of the exciton spot is comparable to the optical wavelength, the relative phases $\phi_n$ have to be taken into account. We may define an overall average Rabi frequency, 
\begin{equation}
  \overline{\Omega_\text R}=\left(\frac{1}{N}\sum\limits_{n=1}^N\,e^{i\phi_n}\right)\Omega_\text R.\label{eq.aveRabi}
\end{equation}
By adjusting a global phase, we assume $\overline{\Omega_\text R}$ to be real and positive. 
For uncorrelated $\phi_n$-values, the emission from the individual excitons will interfere destructively, corresponding to a decrease  of the average coupling $\overline{\Omega_\text R}$ per exciton. On the other hand, if all excitons couple with the same phase $\phi$, $\overline{\Omega_\text R}$ would be independent of $N$ and the excitons emit collectively in a superfluorescent state~\cite{Dicke,Hana-book,Bonifacio,Bonifacio2,Kono}. Superfluorescence is per definition given, if the intensity of the field from $N$ emitters is proportional to $N^2$, see~\cite{Bonifacio}. For no phase matching, on the other hand, the intensity is proportional to $N$.

The effective Hamiltonian may be rewritten as
\begin{equation}
    \hat H/\hbar=\hspace{-0.1cm}\sum_n\hspace{-0.1cm}\left\{\delta\hat a^\dagger_n\hat a_n+\overline{\Omega_\text R}(\hat a_n{+}\hat a^\dagger_n)\right\}{+}G\sum_{n,k}\hat a^\dagger_n\hat a^\dagger_k\hat a_k\hat a_n.
\end{equation}
We can now introduce a transformation, describing the excitons by a single collective bosonic excitation:
\begin{equation}
  \hat A=\tfrac{1}{\sqrt{N}}\sum_n\hat a_n,\quad [\hat A,\hat A^\dagger]=1.
\label{eq.Adef}
\end{equation}
From this, the following commutators can be easily derived:
\begin{align}
  [\hat A,\sum_n\hat a^\dagger_n\hat a_n]&=[\hat A,\hat A^\dagger\hat A],\label{eq.Acom}\\
  [\hat A,\sum_{n,k}\hat a^\dagger_ka^\dagger_n\hat a_k\hat a_n]&=N[\hat A,\hat A^{\dagger2}\hat A^2].\label{eq.Acom2}
\end{align}
The positive frequency part of the emitted source field, $\hat E^{(+)}_\text S$, is proportional to $\hat A$. Thus the correlation properties of the emitted light are characterized by the correlation properties of $\hat A$ and $\hat A^\dagger$. Interpreting the commutators in Eqs.~(\ref{eq.Acom},\ref{eq.Acom2}) accordingly, the source field is correctly described by the collective Hamiltonian
\begin{equation}
	\hat H' =\hbar\delta\hat A^\dagger\hat A+\hbar\Omega_\text R'(\hat A{+}\hat A^\dagger)+\hbar G'\hat A^{\dagger2}\hat A^2\label{eq.S-Ham-2}
\end{equation}
with $\Omega_\text R'=\sqrt{N}\,\,\overline{\Omega_\text R}$ and $G'=NG$. 

In a similar way one can show that, using the general commutation relations
\begin{align}
  [\hat a_n,\hat A^{\dagger m}]&=\tfrac{m}{\sqrt{N}}\hat A^{\dagger m-1},\quad
  [\hat A^\ell,\hat a_n^\dagger]=\tfrac{\ell}{\sqrt{N}}\hat A^{\ell-1},
\end{align}
we find for the Lindblad terms,
\begin{equation}
\begin{split}
  \text{Tr}[\hat A^{\dagger m}\hat A^\ell\mathcal{L}(\hat\varrho)]&=\text{Tr}[\hat A^{\dagger m}\hat A^\ell(2\hat A\hat\varrho\hat A^\dagger{-}\{\hat A^\dagger\hat A,\hat\varrho\})]\\
  &=-(m+\ell)\langle\hat A^{\dagger m}\hat A^\ell\rangle.
\end{split}
\end{equation}
This yields for the total system
\begin{align}
	\dot{\hat\varrho}&=\frac{1}{i\hbar}[\hat H',\hat\varrho]+\frac{\Gamma}{2}\mathcal L'(\hat\varrho)\label{eq.master-2}\\
\mathcal L'(\hat\varrho)&=(2\hat A\hat\varrho\hat A^\dagger-\hat A^\dagger\hat A\hat\varrho-\hat\varrho\hat A^\dagger\hat A),
\end{align}
with $H'$ given in Eq.~(\ref{eq.S-Ham-2}).
Hence, the full cooperative dynamics of our system is identical to the single exciton case -- but with modified coupling constants, depending on the number $N$ of excitons in the spot.
Note that $\Gamma$ and $\delta$ do not scale with $N$. Thus, the effects of multiple excitons only enhance the nonlinear contribution. Broadening of the resulting linewidth is based on many-body effects.

\section{Superfluorescent emission}\label{sec.verify}
The collective Rabi-fre\-quency $\Omega_\text R'$ and the collective nonlinearity $G'$ will be fit parameters of our simulation in sec.~\ref{sec.in-out}. Hence, we use their dependence on $N$ to determine, if the emission of the excitons is superfluorescent. 
The single exciton Rabi-frequency $\Omega_\text R$ also increases with the root of the laser power $P_\text L$. Hence, $\Omega_\text R'/\sqrt{P_\text L}$ gives the $N$-dependence. To analyze the different occurring dependencies, we consider the ratio
\begin{equation}
  \frac{\overline{\Omega_\text R}}{|\Omega_\text R|}=\frac{1}{N}\left|\sum_{n=1}^Ne^{i\phi_n}\right|.\label{eq.phaserel}
\end{equation}
In case of perfect phase matching, the right hand side of Eq.~(\ref{eq.phaserel}) becomes unity, leading to
\begin{equation}
  \Omega_\text R'=\sqrt{N}\,\,\overline{\Omega_\text R}=\sqrt{N}|\Omega_\text R|.\label{eq.collOR}
\end{equation}
Hence, including the power-dependence of $\Omega_\text R'$, and comparing with the $N$-dependence of $G'$, we obtain
\begin{equation}
  \frac{(\Omega_\text R')^2}{P_\text L} \propto G'\propto N.
\label{eq.suprop}
\end{equation}
A more convenient way to describe this result is to compare the parameters for different excitations. Therefore we add an index $i=1,2$ to the quantities above, indicating different fluorescence spectra $1$ and $2$. This yields
\begin{align}
  \frac{(\Omega_{\text R,1}')^2}{P_{\text L,1} G_1'}&=\frac{(\Omega_{\text R,2}')^2}{P_{\text L,2} G'_2},\\
  \frac{P_{\text L,1}}{P_{\text L,2}}\left(\frac{\Omega_{\text R,2}'}{\Omega_{\text R,1}'}\right)^2&=\frac{G_2'}{G_1'}=\frac{N_2}{N_1},\label{eq.superrel}
\end{align}
where $N_i$ are the numbers of excitons involved in the respective spectrum. Note that these numbers depend on the pump power.

From Eq.~(\ref{eq.collOR}), the criterion for superfluorescence follows. The fluorescence intensity $I$ is given by the fields as
\begin{equation}
  I\propto\langle\hat E^{(-)}_\text S\hat E^{(+)}_\text S\rangle=N\langle\hat A^\dagger\hat A\rangle.
\end{equation}
The scaling with $N$ stems from the normalization of $\hat A$ in Eq.~(\ref{eq.Adef}). Using Eq.~(\ref{eq.S-Ham-2}), for weak pumping  we get $\langle\hat A^\dagger\hat A\rangle\propto(\Omega_\text R')^2$. Consequently,
\begin{equation}
  I\propto N(\Omega_\text R')^2=N(\sqrt{N})^2|\Omega_\text R|^2\propto N^2I_1,\label{eq.Isuper}
\end{equation}
where $I_1$ is the single exciton intensity, proportional to $|\Omega_\text R|^2$.

Similar to this derivation, we can estimate the dynamics for a random configuration of excitons without phase matching. In this case we expect destructive interference suppressing a collective dipole coupling of the excitons to the pump field, which yields
\begin{align}
  \left|\sum_{n=1}^Ne^{i\phi_n}\right|^2&=\sum_{n,k=1}^Ne^{i(\phi_n-\phi_k)}\nonumber\\
  	&=\sum_{n=1}^N e^{i\,0}+\sum_{n\neq k}^Ne^{i(\phi_n-\phi_k)}\approx N.\label{eq.sumsolve}
\end{align}
Here we assumed a statistical distribution of the phase differences. This leads to
\begin{align}  	
  	&\left|\sum_{n=1}^Ne^{i\phi_n}\right|\approx\sqrt{N},\\
  	&\Omega_\text R'=\sqrt{N}\,\,\overline{\Omega_\text R}=|\Omega_\text R|,
\end{align}
so that we get in place of Eqs.~(\ref{eq.suprop}),~(\ref{eq.superrel})
\begin{align}
&\frac{(\Omega_\text R')^2}{P_\text L} = \text{constant},\\
&\frac{P_{\text L,1}}{P_{\text L,2}}\left(\frac{\Omega_{\text R,2}'}{\Omega_{\text R,1}'}\right)^2=1.\label{eq.chaosrel}
\end{align}
We have no $N$-dependence of $\Omega_\text R'=|\Omega_\text R|$ for such a random configuration.
Comparing with Eq.~(\ref{eq.Isuper}), we now get
\begin{equation}
  I\propto N|\Omega_\text R|^2\propto NI_1,\label{eq.Ichaos}
\end{equation}
that is, we have only the incoherent increase of the intensity. Hence, we can conclude, that, for a random phase distribution, 
the ratio in Eq.~(\ref{eq.chaosrel}) does not change. On the other hand, an increase of this quantity with increasing pump power indicates superfluorescence, cf. Eq.~(\ref{eq.superrel}).

\section{Emission spectra from quantum wells}\label{sec.in-out}
In order to describe the fields emitted from the quantum well, we will apply the input-output formalism for a light field entering an excited, dispersing and absorbing, but otherwise passive medium~\cite{Dima1}. The results of these methods will be reinterpreted for the case of fluorescence from a quantum film and connected to our above Hamiltonian.
Following the Huttner-Barnett quantization of the Maxwell equations in the presence of dispersing and absorbing bodies~\cite{Huttner,Gruner}, the noise inside the medium is generated by a system of harmonic oscillators, acting as noise sources. These noise sources were introduced in order to conserve the field commutation relation. In~\cite{Dima1} it was shown, that
a light field propagating through an excited semiconductor quantum well consists of a transmitted and a reflected part of the input field $\hat E_\text{in}$, and the spontaneous emission of the induced excitations. The latter, given by the noise operators, were shown to represent the excitons inside the medium. However, in all these cases, the noise operators were merely a bath. In the case of exciton fluorescence, these excitations are driven themselves, so that their Hamiltonian and dynamics must now be substituted with the above master-equation~(\ref{eq.master-2}). Furthermore, as the pumping light is the origin of the excitation of the quantum well, there is no (further) input signal.

The emission spectrum reduces to the spontaneous emission of the medium-induced resonances, which can be given as
\begin{equation}
  S(\omega)\propto a(\omega)\langle\hat {\tilde A}^\dagger(\omega)\hat {\tilde A}(\omega)\rangle.
\label{eq:fullspectrum}
\end{equation}
Herein, the operators $\hat {\tilde A}(\omega)$ and $\hat {\tilde A}^\dagger(\omega)$ represent the Fourier-transformed annihilation and creation operators of the induced resonances of the medium -- the excitons and polaritons. Hence, $\langle\hat {\tilde A}^\dagger(\omega)\hat {\tilde A}(\omega)\rangle$ is identified as the emission spectrum of these quasiparticles. The frequency dependent function $a(\omega)$ describes the absorption of the material and can be computed as
\begin{equation}
  a(\omega)=1-|t(\omega)|^2-|r(\omega)|^2,
\end{equation}
with $t(\omega)$ and $r(\omega)$ being the complex transmission and reflection coefficients of the medium, respectively.

For quasi-equilibrium, the emission spectrum is supposed to approach a Bose distribution~\cite{Henneb1}, which is in accordance with the above discussion on the noise operators of Huttner and Barnett. As absorption and emission are both accessible to experiments, it was proposed to test an excited semiconductor for quasi-equilibrium. However, such experiments, performed on ZnSe~\cite{Seemann}, indicated ``that the detected emission comes from the polariton states excited by the pump pulse, rather than from a quasi-equilibrium distribution.''
Consequently, we apply the different interpretation of using the dynamics from Eq.~(\ref{eq.master-2}). The emission spectrum of the quasiparticles is simply the Wiener-Khintchine spectrum, 
\begin{equation}
	S_\text{W} (\omega) \propto\hspace{-0.15cm} \int\limits_{-\infty}^{\infty}\hspace{-0.15cm}d\tau\, e^{-i\omega\tau}\lim_{t\rightarrow\infty}\langle\hat A^\dagger(t)\hat A(t+\tau)\rangle\propto \langle\hat {\tilde A}^\dagger(\omega)\hat {\tilde A}(\omega)\rangle,\label{eq.identity_b_S}
\end{equation}
obtained from the Fourier-transform of the two-time correlation function.

The Wiener-Khintchine spectrum can be calculated from our collective bosonic operators $\hat A$ via the quantum regression theorem, see e.g.~\cite{vowe}. Thus we get the full emission spectrum of the quantum well, $S(\omega)$, from Eq.~(\ref{eq:fullspectrum}). It combines the excitonic emission spectrum inside the well, $S_\text{W}(\omega)$, with the many-body effects included in the absorption spectrum $a(\omega)$ of the passive medium. The absorption will be calculated from transmission and reflection as given in~\cite{Dima1}, with an oscillator-model for the susceptibility $\chi(\omega)$:
\begin{equation}
  \chi(\omega)=\frac{f}{\omega-\omega_\text X-i\tfrac{\Gamma}{2}}=\frac{f}{\omega-\omega_\text L-\delta-i\tfrac{\Gamma}{2}}.
\end{equation}
The width $\Gamma$ and resonance frequency $\omega_\text X$ of this oscillator model are equal to the exciton parameters in Eq.~(\ref{eq.Master}).

\begin{figure}[h]
  \includegraphics[width=8cm]{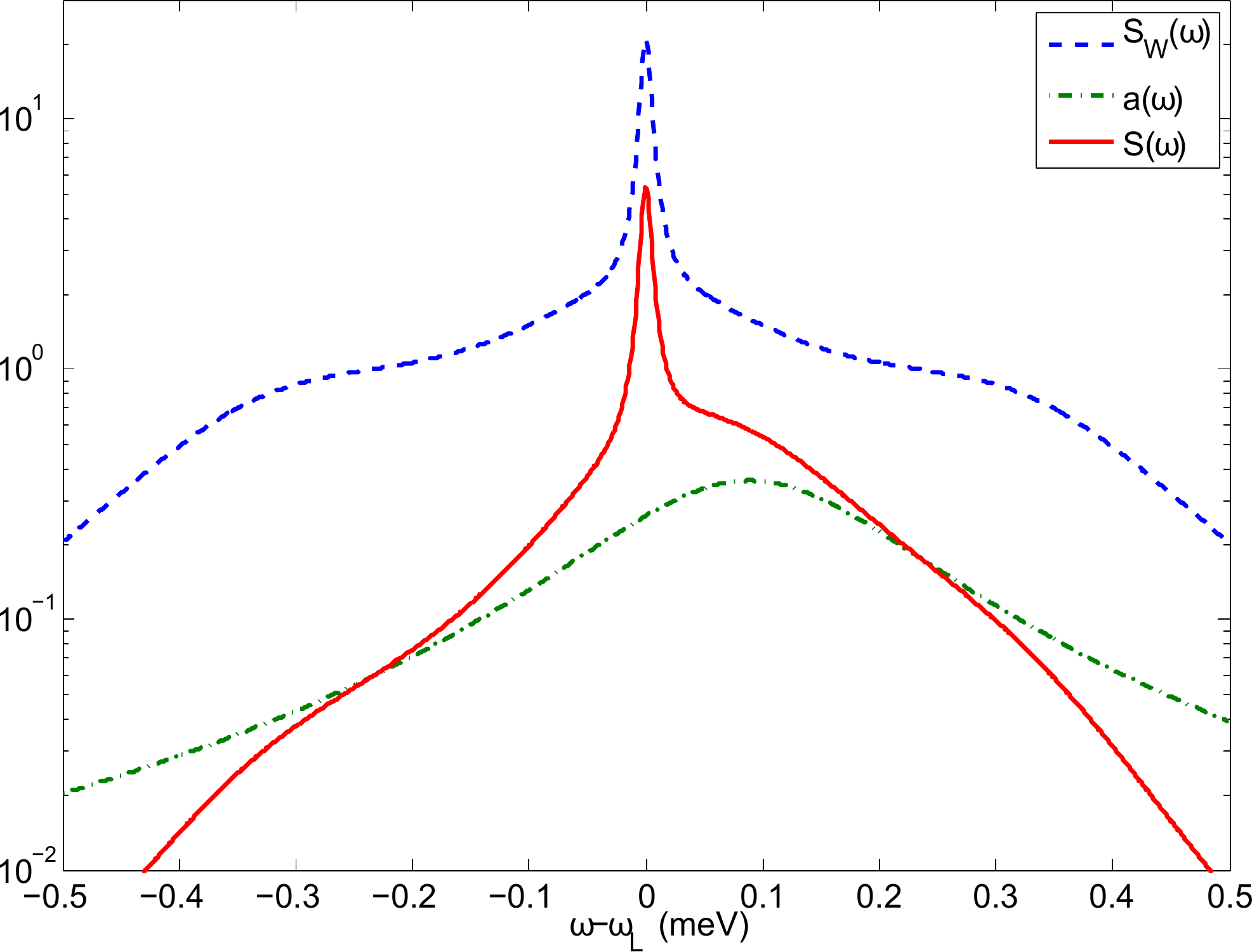}
\caption{(color online.) Theoretical emission spectrum of the excitons inside (dashed) and outside the quantum well (solid). The dash-dotted line gives the absorption spectrum. The parameters are $\Gamma=0.22$ meV, $G=0.45$ meV, $\Omega_\text R=0.16$ meV, $\delta=0.1$ meV and the detector bandwidth $\Gamma_\text f=0.0107$ meV.}\label{fig.specTheo}
\end{figure}

In Fig.~\ref{fig.specTheo} we compare the theoretical spectrum inside and outside the quantum well for the parameters as given in the caption. The width of the Rayleigh-peak stems from the limited resolution of the detector, modeled by a Lorentzian of width $\Gamma_\text f$. The spectrum inside the well is symmetric with respect to the laser frequency as expected from energy conservation. The incoherent emission and thus the sidebands of the exciton spectrum are due to the Kerr-nonlinearity, which would  produce Rabi-splitting for larger values of $G'$ or $\delta$, see~\cite{Liu}.
Note that the resonance of the absorption is shifted from the laser resonance. For thin films, whose width is small compared to the wavelength of the absorption resonance, $a(\omega)$ is proportional to the imaginary part of $\chi(\omega)$. Consequently, the emitted spectrum $S(\omega)$ appears asymmetric outside the well. Furthermore, its resonance frequency (maximum of the incoherent part of the emission) is shifted in the same direction as the absorption, but to a significantly smaller degree. This shift was observed in the experiments~\cite{Burau-1}.

\begin{figure}[h]
 \includegraphics[width=7.5cm]{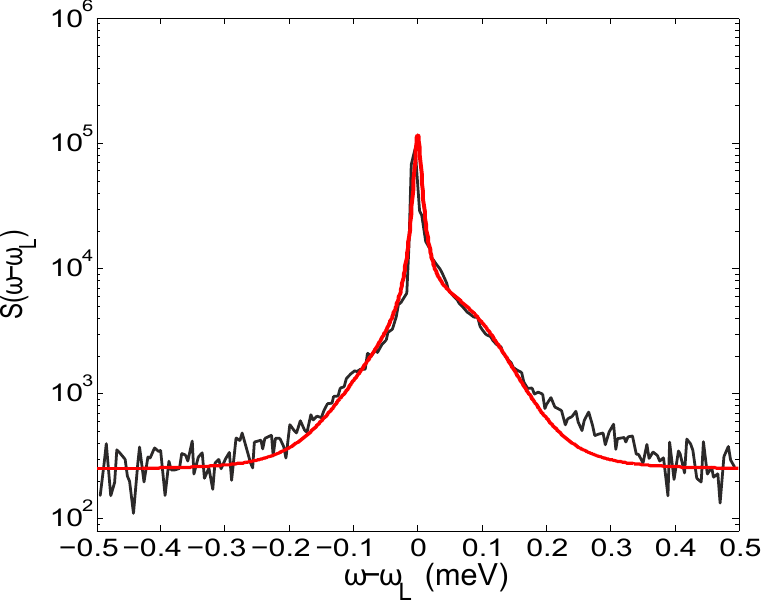}
  \includegraphics[width=7.5cm]{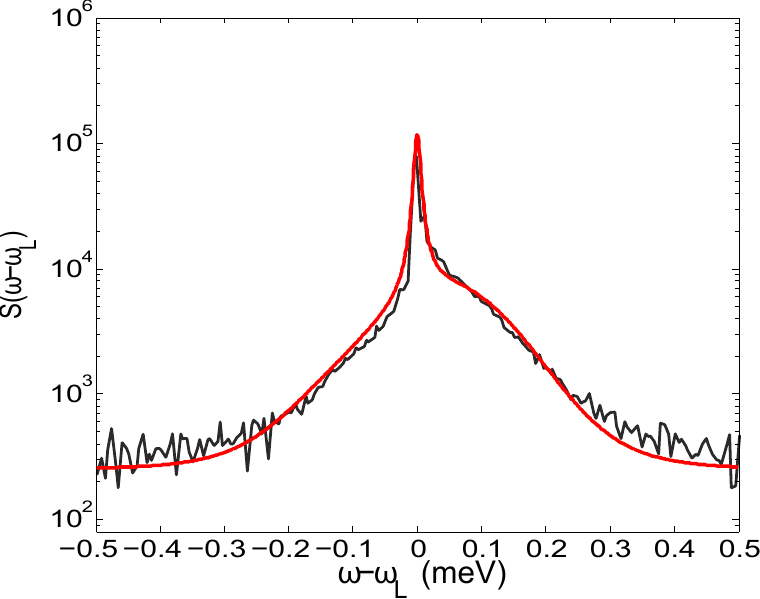}
  \includegraphics[width=7.5cm]{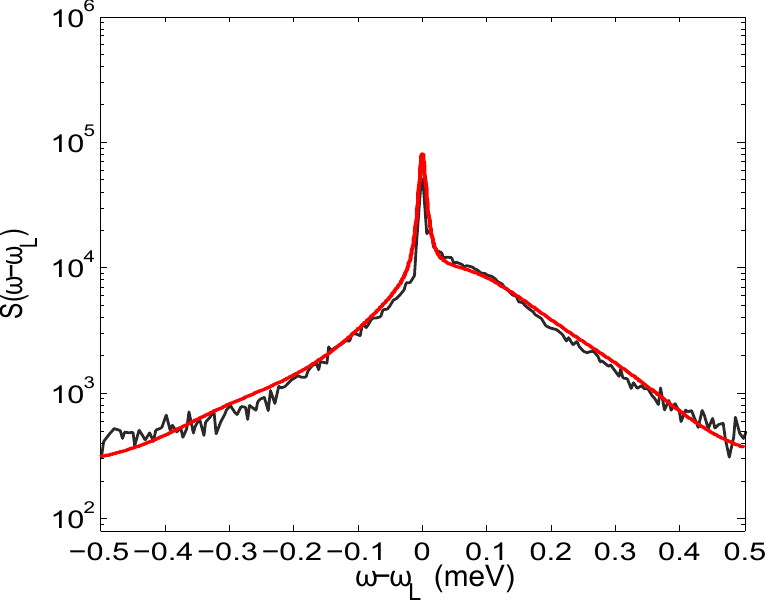}
  \caption{(color online.) Spectra of excitons, comparison of theory (red, bright) with experiment (black, dark) for different laser powers $P_\text L$. Top: $P_\text L=100$ $\mu$W, middle: $P_\text L= 150$ $\mu$W, bottom: $P_\text L= 310$ $\mu$W.}\label{fig.spectra}
\end{figure}

In Fig.~\ref{fig.spectra} we compare the experiments for the three laser powers $P_\text L=$ 100, 150 and 310 $\mu$W, also cf. Fig.~\ref{fig.ExpRF}, with our theoretical model and obtain very good agreement for the spectra.  The fit parameters are given in the table~\ref{tab.fit}. The Rayleigh peak itself is symmetric, since it is very narrow compared with the width of the absorption spectrum $a(\omega)$: $\Gamma_\text f=0.0107$ meV for all powers, while $\Gamma\geq0.15$ meV. 
The apparent shift of the quantum well spectrum  is due to the shift of the absorption resonance. However, the slow increase of the detuning $\delta$ with increasing laser powers yields two conclusions: On one hand the majority of the shift is based on the detuning of the laser from the absorption resonance, not on many-body effects.
On the other hand the density of excitons is much lower than estimated in~\cite{Burau-3}. The latter statement is supported by the comparably small increase of $\Gamma$ and decrease of $f$,  indicating that many-body effects are small in this regime. In order to account for the detection noise, appearing in the experiment, we supplemented our theoretical spectra with a constant background. Due to this noise and the fact that the exciton spectrum $S_\text W(\omega)$ is multiplied with $a(\omega)$, physical effects outside the absorption resonance are suppressed.

\begin{table}[h]
   \begin{tabular}{|c||c|c|c|}\hline
    measurement $i$ & 1 & 2 & 3\\
      \hline
    $P_{\text L,i}$/$\mu$W & 100 & 150 & 310\\
      \hline\hline
    $\hbar G'_i$/meV & 0.10& 0.205& 0.45\\
      \hline
    $\hbar\Omega_{\text R,i}'$/meV & 0.045& 0.075& 0.16\\
      \hline
    $\hbar\delta_i$/meV & 0.08& 0.08& 0.09\\
      \hline
    $f_i$/a.u. & 1.0& 1.0& 0.9\\
      \hline
    $\hbar\Gamma_i$/meV & 0.15& 0.20& 0.22\\
      \hline
    $G_{i}'/G_{i-1}'$ & $-$ & 2.050 & 2.195\\
      \hline
    $\frac{P_{\text L,i-1}}{P_{\text L,i}}\left(\frac{\Omega_{\text R,i}'}{\Omega_{\text R,i-1}'}\right)^2$ & $-$ & 1.852& 2.202\\
      \hline
  \end{tabular}
\caption{Comparison of the different fit parameters for the different measured spectra. 
}\label{tab.fit}
\end{table}

The decrease of the Rayleigh-peak while the pump laser intensity is increased can also be explained. The amplitude of the Rayleigh-peak is proportional to the coherent part of the collective excitation $|\langle\hat A\rangle|^2$. Taking into account the collective increase of the Kerr-nonlinearity $G'$ and Rabi-frequency $\Omega_\text R'$, this coherence should always increase with increasing $P_\text L$.
However, in addition to these effects, the quantum well emission fields are diminished by the absorption. As the exciton resonance shifts out of the laser resonance and broadens due to many-body effects, the absorption at the laser frequency, $a(\omega_\text L)$, reduces substantially. Likewise, the coherence is decreasing for increasing  detuning $\delta$ and decay rate $\Gamma$. Both effects combined overcompensate the expected increase of the coherence and thus the amplitude of the Rayleigh peak.

\section{Signature of Superfluorescence} \label{sec.super}

For the purpose of analyzing the experimental spectra, we consider the parameters of our theory which yield the best fit of the experiments, see table~\ref{tab.fit}. To compare with our calculations above concerning the relative increase of $G'$ and $\Omega_\text R'$, we added the index $i$ to all quantities, to number the respective measurement, compare Sec.~\ref{sec.verify}. The ordering of the spectra is by increasing laser power $P_\text L$, so that $P_{\text L,i+1}>P_{\text L,i}$. The last two lines of the table are used for the comparisons. The first of them gives the relative increase of $G'$, while the last one gives the left hand side of Eq.~(\ref{eq.superrel}).

The increase of $G'$ indicates that the number $N$ of excitons involved in the exciton spot roughly doubles for each increase of the laser excitation. From our considerations we conclude, that the last line of the table should yield the same value as the increase of $G'$ for the case of superfluorescence, but remains equal to one for random phases. Intermediate values indicate a partial phase matching.
As can be seen from table~\ref{tab.fit}, the values in the last line significantly exceed the value of one. In particular, from the second to the third measurement the value is very close to the relative increase of $G'$, in very 
good agreement with Eq.~(\ref{eq.superrel}). Thus we conclude, that the phase matching is fairly good, and the excitons do emit steady-state superfluorescent light.

\section{Conclusions and Outlook}\label{sec.con}
We have studied the resonance fluorescence of a localized exciton system in a quantum well experimentally and theoretically. In the experiments, despite localization, the excitons did not show vacuum-Rabi splitting as known for single-photon emitters in strong light-matter coupling. Based on these results we formulated a low-excitation Hamiltonian of interacting bosonic excitons. We showed that the excitons behave like a single exciton with collective parameters. Combining the quantum optical description of the exciton emission spectra with methods for analyzing the propagation of quantum fields in media, we obtained the fluorescence spectra of laser-driven quantum wells, matching very well the experiments. These results allowed us to clarify the discrepancy in the magnitude of the pumping induced shift of the exciton resonance, as well as the decrease of the Rayleigh peak with increasing strength of pumping. Furthermore, we could show, that the excitons do collectively emit steady-state superfluorescent 
light. Based on our model, we can describe the relative increase of the number of excitons within one exciton spot. The present theory may be useful to predict results of other experimental scenarios, such as the general quantum correlation properties of excitonic light sources in semiconductor structures.

\acknowledgments
This work was supported by the DFG through SFB~652.

\end{document}